\pdfoutput=1
\PassOptionsToPackage{table,xcdraw}{xcolor}
\documentclass[letterpaper,twocolumn,10pt]{article}
\usepackage{usenix2019_v3}
\usepackage{textgreek}
\usepackage[cachedir=_minted-paper,frozencache=true]{minted}
\usepackage{dirtree}
\usepackage{times}
\usepackage{amsmath}
\usepackage{adjustbox}    
\usepackage{booktabs}
\usepackage{graphicx}
\usepackage{listings}
\usepackage{subcaption}
\usepackage{framed}
\usepackage{soul}
\usepackage{textcomp}     
\usepackage{tikz}
\usetikzlibrary{positioning}
\usepackage{xspace}
\usepackage{tabularx}
\usepackage{amssymb}
\usepackage{multirow}

\usepackage[capitalise,nameinlink,noabbrev]{cleveref}     
\usepackage{enumitem}

\def \ie {\textit{i.e.,}\xspace}
\def \eg {\textit{e.g.,}\xspace}

\def \snapfaasnoprof {$\name{}-$\xspace}
\def \snapfaas {$\name{}$\xspace}
\def \seuss {$SEUSS_{SF}$\xspace}
\def \reap {$REAP_{SF}$\xspace}

\def \regular {\texttt{regular}\xspace}

\def \rootfs {\texttt{RootFS}\xspace}
\def \appfs {\texttt{AppFS}\xspace}

\def \base {\texttt{base}\xspace}
\def \diff {\texttt{diff}\xspace}

\makeatletter
\renewcommand{\paragraph}{%
  \@startsection{paragraph}{4}%
  {\z@}{1.75ex \@plus 1ex \@minus .2ex}{-1em}%
  {\normalfont\normalsize\bfseries}%
}
\makeatother

\begin{document}
\newcommand{\name}{SnapFaaS}
\date{}

\title{\Large \bf How Low Can You Go?\\ Practical cold-start performance limits in FaaS}
\author{Yue Tan, David Liu, Nanqinqin Li, Amit Levy\\
Princeton University}


\maketitle
\thispagestyle{empty}
\begin{abstract} 
Function-as-a-Service (FaaS) has recently emerged as a new cloud computing paradigm.
It promises high utilization of data center resources through allocating
resources on demand at per-function request granularity.
High cold-start overheads, however, have been limiting FaaS systems' such potential.

Prior work has recognized that time redundancy exists across different cold function
invocations and has proposed varied snapshots that capture
the instantaneous execution state so allow for jump-starts through restoration.
However, it remains unclear what the cold-start performance limits are as the previous
snapshots employ different techniques and are designed for different environments.
In this paper, we summarize these snapshots from a taxonomic perspective and present
a model that depicts the cold-start performance from first principles. To
approximate the performance limits, we propose a snapshot design \name{}.
We evaluate \name{} using real-world FaaS functions. Our empirical results prove
\name{}' efficiency. It is 2-10x as fast as prior work for most
functions and achieves near-optimal cold-start performance.
\end{abstract}

\section{Introduction}\label{sec:introduction}

Function-as-a-Service (FaaS) is a cloud computing paradigm where developers
write containerized ``functions'' which the cloud platform runs and invokes on
demand in response to end-user requests, storage events, and other platform
events~\cite{aws-lambda, cloudflare-workers, google-functions,
azure-functions}. Through fine-grained resource allocation and the flexibility
to free resources once a function invocation completes, FaaS promises high
utilization of data center resources, even when workloads are heavy-tailed.

In practice, this promise is encumbered by the resource and performance
overhead of initializing the system abstraction that FaaS workloads rely on: a
full Linux environment and high-level language runtime.

FaaS systems typically allow developers to write functions in one of a few high
level languages with access to a complete, though often stripped down, Linux
distribution. This affords developers the flexibility to leverage myriad
existing native libraries and programs available for Linux as well as the
richness of a full operating system (e.g.\ a POSIX-shell, performant TCP/IP
implementations, etc).

Unfortunately, initializing such an environment is costly compared to the
typical FaaS function runtime. While most functions execute for less than one
second~\cite{serverless-in-the-wild}, initializing the environment for a
function (initializing a virtual machine and kernel or container, running OS
init scripts, starting a language runtime, and importing libraries) takes 100s
of milliseconds on recent production hardware~\cite{firecracker-nsdi}. This is
a problem for both end-to-end request latency and utilization. Long
initialization times significantly lengthen the otherwise short end-to-end
request latency and tie up CPU and memory that could have been used  respond to
other requests.

Researchers in the past few years have termed ``cold-starts'' the problem of
initializing FaaS functions.  Practical systems mitigate this overhead for
popular functions by keeping recently executed functions ``warm'' for a
period---simply keeping the virtual machine or container running---in hopes
that another request for the same function arrives soon~\cite{gcf-warmup-doc,
firecracker-nsdi, serverless-in-the-wild}. This can improve end-to-end request
latency for frequently requested functions. However, a large portion of
requests are to functions invoked less than once a
minute~\cite{serverless-in-the-wild} and keeping all such functions warm
sufficiently long is an impractical use of CPU and memory resources. As a
result, keeping functions warm, while useful, only improves end-to-end request
latency for popular functions and wastes resources when functions kept warm are
\emph{not} invoked again.

A flurry of research addresses this problem by trying to improves function
cold-start themselves.  In particular, much prior work has correctly identified
that initialization is almost identical for different invocations of the same
function. Moreover, much function initialization is redundant across different
but similar functions since they often share a kernel, OS init system, and run
atop one of a small number of language runtimes. Notably, most work uses some
form of memoization to bypass portions of initialization~\cite{seuss, sock,
catalyzer, replayable-execution, reap}---replacing logical initialization code
with load memory directly from disk.

These research systems reduce the end-to-end latency to handle a cold request
for a noop function from nearly one second to around 100ms. However, prior
systems target different environments and use different memoization techniques,
some of which are complimentary while others conflict. It remains unclear what
the lower-bound cold-start is, and whether prior approaches are sufficient to
achieve performance such a lower-bound.

In this paper, we propose a principled framework for identifying the pracical
limits of cold-start performance and describe \name{}, a VM snapshot based FaaS
system that achieves significantly better cold-start performance than prior
work and nears the practical lower-bound.

We start by defining cold-starts and function memoization, and presenting a
taxonomy of prior approaches (Section~\ref{sec:background}).  We then identify
the minimal necessary state to handle a function request
(Section~\ref{sec:model:state}) under practical resource constraints, and offer
a first-principles model that characterizes cold-start performance
(Section~\ref{sec:model:model}).

We then describe \name{} the design (Section~\ref{sec:design}) and
implementation (Section~\ref{sec:design}) of \name{}, including a VM snapshot
technique guided by the model. \name{} uses fixed memory overhead to cache
shared partial snapshots in memory but loads all function-specific memoized
state from disk. Key to achieving this is coupling the virtual block and
network device configuration, software initialization steps, and snapshot
orchestration to ensure that the most sharable memory is initialized before
\emph{any} function-specific state.

We evaluate \name{} and show that it always outperforms end-to-end request
latency in prior systems---typically by 2-10x. Moreover, \name{} achieves near
optimal end-to-end latency overhead for cold requests (Section~\ref{sec:eval}).
We conclude by discussing what impacts these results should have on future work
in the area.


\section{Cold-Start Mitigations in FaaS Today}\label{sec:background}
In FaaS, physical machines multiplex executions of functions by running each in
a virtual machine or OS container and allocating fixed resources (CPU, memory,
network, and ephemeral disk) to each function while it executes or sits idle.
Typically, CPU and memory are the limiting resource and, indeed, public FaaS
deployments typically bill in CPU-memory time units.

Function requests are often cold---meaning there is no running container or VM
available running the function available to service the request. Production
systems often keep function instances ``warm'' (i.e.\ they do not kill the VM
or container) for a grace period even if no requests for the function arrive to
improve latency for popular functions~\cite{serverless-in-the-wild}. In this
work we only consider cold-starts, though such strategies for optimizing
frequently invoked functions are complementary when there is an abundance of
memory resources.

We also only consider end-to-end request latency since, unlike in interactive
systems, booting a function quickly at the expense of a delaying the final
response by the same amount is rarely useful in FaaS.

\subsection{Reducing Cold-Starts with Memoization}\label{sec:background:snapshot}
FaaS functions are written in high level languages atop managed runtimes such as
Python, JavaScript, Java, Ruby, and C\#. Functions also use native libraries
and executables that leverage a UNIX-like system interface and standard UNIX
shell utilities (\eg \texttt{ls}, \texttt{echo}, etc). Most commonly, FaaS
platforms expose a complete Linux-based OS environment to
functions. For example, a function that creates an image thumbnail has a
handler function written in Python that fetches the original image from an
object storage service (\eg S3), invokes a shell command to run the
ImageMagick \texttt{convert} command-line utility to generate a thumbnail and,
finally, uploads the result back to the object storage service.

In order to service a request, the OS, language runtime, libraries, etc. must
be initialized to a state that is able to service it---namely to invoke the
function's entry point \texttt{handle} procedure. Initializing the kernel, OS,
language runtime, libraries and function initialization code generates this
state, by definition. However, it is not \emph{necessary} to run this code each
time, even though each component may be non-deterministic.

Specifically, FaaS platforms require functions to be written such that
\emph{any} state that was the result of executing the above initialization
steps is sufficient to correctly invoke the function's \texttt{handle}
procedure, as long as certain general invariants are maintained---the real time
clock should be correct, the network should be functional, etc. This makes it
possible to memoize much of the logic performed during cold start and replace
relatively slow initialization execution with relatively fast data copying.


\subsection{Prior Work}\label{sec:background:priorwork}
Prior work has proposed varied memoization techniques suited for different environments to
improve cold-starts.  Catalyzer~\cite{catalyzer} is designed for
gVisor~\cite{gvisor}, SOCK~\cite{sock} is designed for Docker~\cite{docker},
SEUSS is designed for unikernel~\cite{unikernel} virtual machine (VM), REAP is
designed for the Firecracker microVM~\cite{firecracker-nsdi}.

Despite different runtime environments and technical details, these techniques
share the same high-level ideas---memoization should capture as much initialization
computation as possible, and the restoration should reduce the amount of
state restored from disk.

One straightforward form of memoization, a snapshot taken after the function is
initialized (a full-function snapshot), captures an initialized but not yet
invoked function's execution state as a whole.  Catalyzer's func-image is a
such. A func-image is generated through conventional container checkpoint
mechanism. It contains the guest's (the function and its runtime environment)
memory and metadata (host-side state).  Booting from a func-image reconstructs
the guest's address space through demand-paging using file-\texttt{mmap}.
On-demand restoration avoids prefetching the whole state from disk but pays for
synchronous page faults at runtime.

REAP recognizes that synchronous page faults incur high runtime latency penalty
and proposes an optimization that prefetches only an approximated working set.
Specifically, after taking a full-function snapshot where the entire memory of
a pre-initialized function is captured, REAP runs the function once, observing which
pages from the snapshot are actually accessed during execution. On future executions,
REAP prefetches only that subset of memory, leaving the rest on disk to be demand-paged.
The in-batch prefetching significantly reduces total latency overhead through saving
considerably many synchronous page faults to disk at runtime.

The other systems take a different route. They cache partial
state that is common and therefore sharable to many functions in memory. Typically, such partial
state results from up to language runtime initialization but no function initialization.


Catalyzer, to further improve cold-starts, proposes language template
Zygote~\cite{zygote}. A Zygote is an idle container having completed some
initialization from which a new container can be spawned and specialized. For
example, any Python function can be booted from the same Python Zygote.
Functions are initialized from Zygotes using the \texttt{fork} system call to
create a copy-on-write clone of the Zygote and loads function code in the
clone.


SOCK, similar to Catalyzer, uses Zygotes that have certain Python packages imported.
Its goal is to save package importing times. SOCK also relies on \texttt{fork}
syscall to spawn new containers but it proposes a different protocol suited for
Docker.

SEUSS uses VM based runtime snapshots cold requests. A runtime snapshot
includes the VM's physical memory from the moment the language runtime
initialization completes. Any function in the language can be booted from the
same runtime snapshot. Function initialization then starts from the restored
runtime. SEUSS uses \texttt{mmap} syscall to implement copy-on-write semantics.

We note that when separating out and caching common state, the designs above
fail to memoize function initialization as full-function snapshots do.  In
fact, SEUSS and Catalyzer, each additionally proposes in-memory
``full-function'' snapshots of some form. Catalyzer has function template
Zygotes and SEUSS has function snapshots. Function template Zygotes are like
the language ones but captures function initialization as well. Function
snapshots, enhancing runtime snapshots, capture any memory modified during
function initialization starting from the correct runtime snapshot.

Caching ``full-function'' state in memory is fast and can help burst
scalability when a function instance is already running anyway. However, each
stored state consumes memory proportional to the number of functions so is
inappropriate for speeding up cold-starts for all functions that might be
invoked.

\section{Memoization from First Principles}\label{sec:model}

Existing techniques use similar insights---minimize initialization work by
memoizing function state and minimize restoration from disk. However, the
techniques are designed for different systems (VMs, containers, unikernels,
etc) with inconsisent views of system constraints, such as how much memory is
permissible to ``waste'' on caching states in memory.

We begin addressing the cold start problem by starting from first-principles.
What is the minimal work that dictates how fast function memoization can be?
How do scalability and resource constraints dictate where such states must be
stored? Finally, what is the least restoration work an idealized system should
do?

\subsection{Required Execution State}\label{sec:model:state}
We start by looking at the state sufficient and necessary to service a request,
and we use Firecracker microVM as a concrete example from now on.
The state necessary to service a request includes:

\paragraph{CPU Registers.} Each CPU core allocated to a function instance has a
few dozen unique word-sized registers (\eg the stack pointer \textit{rsp} and
the instruction pointer \textit{rip} on x86\_64) required to run.

\paragraph{Virtual Device State.} Virtual devices (\eg VirtIO~\cite{virtio} block and
network devices in a hypervisor) are state machines with relatively
simple states: a few pointers to function memory for device-VM shared buffers and device
specific state, such as the MAC address of a virtual network device.

\paragraph{Initialized \& Useful Memory.} Executing a function from its entry point
procedure relies on certain parts of the memory being resident in the main memory---
all memory pages that are written to during environment initialization and actually
used during execution. This constitutes the largest portion of the program state---
up to 54~MB in our experiments.

These parts include memory pages modified by the kernel, OS init process, the
language runtime, any base libraries used by the language or FaaS runtime
(typically a language-specific library), and function initialization. Some of
these memory pages are common across similar functions (\eg those that use the
same kernel, OS userland and language runtime) and some are specific to each
function.

\subsection{Fundamental Overheads}\label{sec:model:model}
Snapshots boost cold-starts by memoizing initialization and turning cold-starts
initialization-less.  However, restoring a memoized function is not
instantaneous as copying data takes time as well.

In general, there are two options for restoring state from disk.  First, state
can be restored on demand, where as the function attempts to access missing
memory pages, the hypervisor loads the page, synchronously, from disk. As a
result, only useful memory pages are loaded and the function can begin
executing virtually instantly. However, demand-paging is synchronous---the
function blocks until each page is fetched into the main memory---pushing the
overhead of restoration to function execution time and preventing batching.
Conversely, eager restoration loads memory pages from disk ahead of function
execution, \emph{in batch}. This delays the start of a function and may load
pages that are never used, but runs at the storage medium's bandwidth speed.

In the best case, snapshots are themselves cached in memory and restored on demand.
Such caching must be done in a sophiscated way. Because there
are many more functions that \emph{could} possibly run than could fit in memory
on each machine, any state proportional to the number of functions cannot be
stored long-term in memory. As a result, any function-generic state, as it can be
shared across functions through copy-on-write semantics, can reside in memory.

The higher latency the storage medium, the more important it is that snapshot
memory be loaded eagerly. While disk is both lower bandwidth and higher
latency than memory, latency is a much more significant factor. For example,
modern RAM has about 50x the bandwidth of an SSD (~500Gbps compared to
~10Gbps), RAM latency is 5 orders of magnitude faster than SSD read latency
(~100ns compared to ~16us).

As a result, CPU-registers, virtual device state, and function-specific memory
should reside on disk and be restored eagerly while function-generic
memory should reside in memory and be restored on demand.

In summary, to reconstruct the state of a VM prepared to handle a request, we must:

\begin{enumerate}
  \item load CPU registers and restore virtual device state ($c$)
  \item Eagerly restore from disk non-zero memory pages unique to each function
    ($pgs_{unique}$)
  \item execute any remaining initialization code who's resulting state
    cannot be captured by a snapshot (e.g.\ reading and deserializing the
    request payload) ($init$)
  \item And copy from memory any snapshot pages that can be shared amongst
    functions that are written to during function execution
    ($pgs_{shared}$)
\end{enumerate}

Only the first two steps can occur concurrently. CPU registers, virtual
device state, and any memory not paged on demand must be available
before any remaining initialization which, in turn, must run before the
function begins executing.

Therefore, given the memory latency $lat_{mem}$, disk bandwidth $bw_{disk}$,
and page size $P$ (typically 4KB), the minimum overhead to end-to-end
performance that \emph{must} be incurred by any snapshot restoration strategy
is given by:

\begin{equation}
  max(c, (\frac{pgs_{unique}\times P}{bw_{disk}})) + init + (pgs_{shared}\times lat_{mem})
\end{equation}

We can see by the model that to achieve the best cold-start performance is to
minimize each of these terms by: splitting snapshots such that the most pages
are sharable amongst many functions but are seldom written to; identifying the
minimum number of unique pages that must actually be restored for each
function; and minimizing the amount of initialization code necessary for each
function invocation.

\section{\name{}}\label{sec:design}
Following the goals above, we propose SnapFaaS a snapshot
based on Firecracker microVM that testifies the cold-start performance limits.

At a high-level, the design consists of one in-memory \base snapshot for each language
runtime and one \diff snapshot and one working set (WS) file, both on-disk, for each function.
There are two key techniques accompanying the design---1) use of two file systems to allow
easy \base/\diff separation and 2) coupled guest-host network configuration.


\subsection{Maximize Shared Pages, Minimize Shared Written Pages}

The first goal is to maximize the number of in-memory sharable pages (i.e.
minimize $pgs_{unique}$) and minimize the ratio of these sharable pages that
are written to during execution. \name{} accomplishes this by generating
separate snapshots: a \base snapshot for the common ``runtime'' and a \diff snapshot
for a function and the libraries it imports.

The \base snapshot includes memory initialized by the kernel, OS, language
runtime, base libraries, and the \name{} runtime. Because all functions use one
of a small number of language-runtimes, each such \base snapshot can be shared
across many functions. The \diff snapshot includes memory initialized or modified
by the function itself and the libraries it imports. This may include memory pages
that were also initialized in the \base snapshot.
In this case, \diff snapshot values override \base snapshot values.

\begin{figure}[t]
  \centering
  \includegraphics[width=\linewidth]{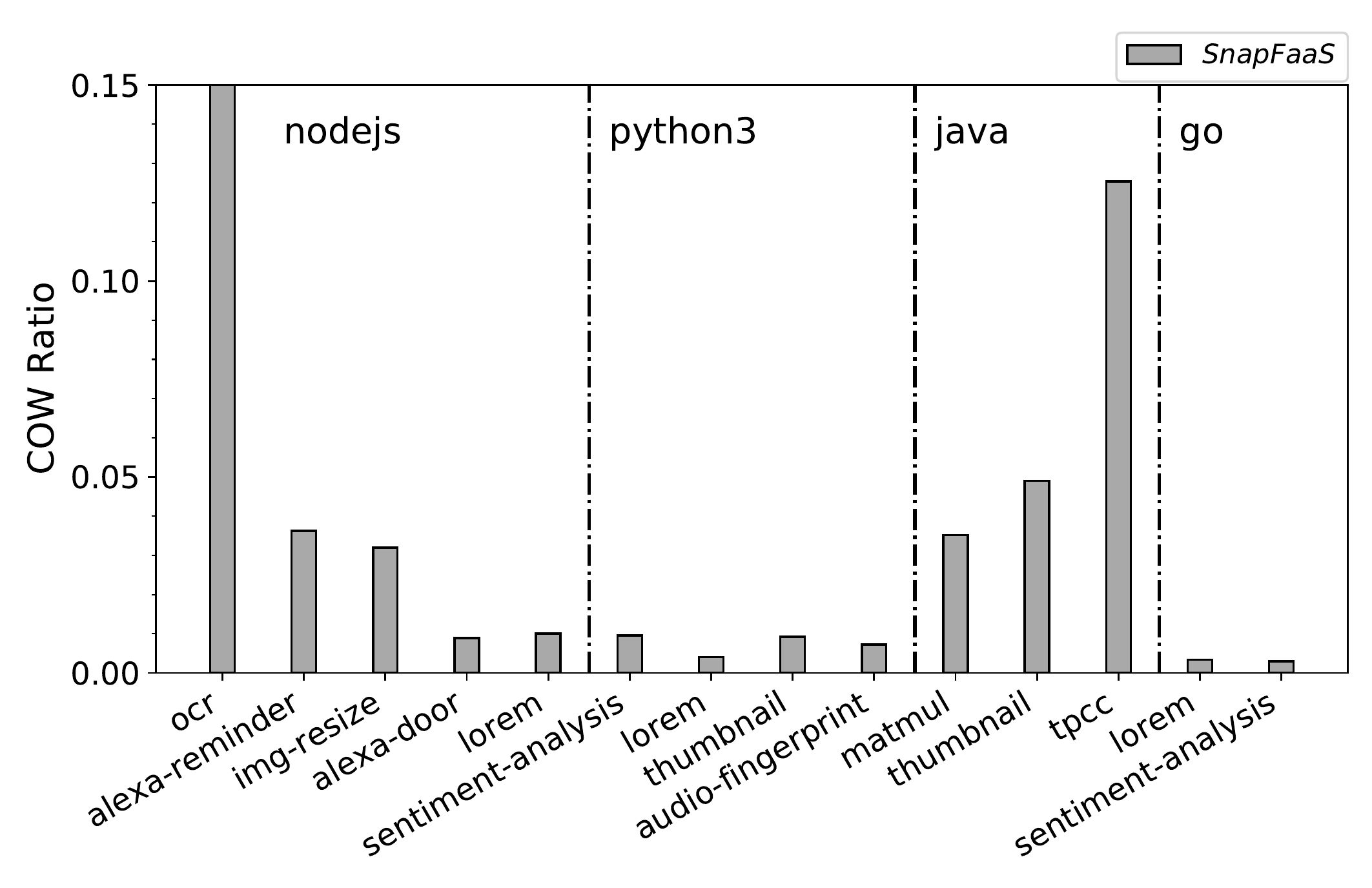}
  \caption{Copy-on-write ratio. The ratio is calculated as the number of
  memory pages that are modified by the language runtime initialization
  and are written to during execution over the total number of
  shared in-memory pages. The numbers are from a single run.}
  \label{fig:cow-ratio}
\end{figure}

While functions may use any of this \base snapshot during execution, each
function uses a relatively small portion of this memory. Moreover, as shown in
Figure~\ref{fig:cow-ratio}, functions \emph{write} to very few of these shared
memory pages during execution. At most 15\%, and more typically fewer than 5\%,
of pages present in the base snapshot are written during function execution.
The result is that $pg_{shared}$ is small and, consequently, few pages are copied on demand.

\paragraph{Key technique---use of \appfs.}
In order to facilitate layering of \diff snapshots on top of \base snapshots, a
\name{} VM uses an application file system (\appfs) in addition to a must root file system
(\rootfs) that stores boot-critical programs like \texttt{/init} 
and OS \& language utilities.

Functions can be packed into the \rootfs as well. However, it will be much trickier
to ensure that \base snapshots do not include the result of executing any
function-specific initialization. This is because once a block device is mounted,
file system metadata, such as the root inode, are cached in memory by the Linux
kernel and, as a result, a snapshot captures specific layout of the file system.

Use of \appfs solves the layout problem because the \appfs is not mounted when
the \base snapshot is generated. Additionally, the separation easily realizes
one root file system for each supported language.
\name{} currently provides four \rootfs-es, Python3, Node.js, Java, and Go.

\subsection{Minimize Unique Pages}

The second goal is to identify the minimum set of unique pages that must be
actually resident in memory. To achieve this goal, \name{} draws on REAP's
technique~\cite{reap}.

Similar to REAP, \name{} approximates a function's working set using the
working set from one previous execution. Then \name{} eagerly reads in the
working set and demand-paging the rest.  Note that \name{} only applies this
technique to \diff snapshots since the purpose is to minimize the number of
pages fetched from disk into memory.

\subsection{Minimize Initialization Computation}

\name{} captures nearly all initialization code in the \base and \diff
snapshots. Capturing some of this is trivial. For example, initialization in
libraries that sets up data structures, loads an ML model from disk, etc, are
independent of invariants outside the function VM, so are captured entirely by the
memory encoded in the \diff snapshot. However, some other initialization code
captures state \emph{outside} the VM and, therefore, requires tight coupling
with the FaaS platform.

\paragraph{Key technique---coupled guest-host network configuration.}
For example, the TCP/IP stack in Linux has internal states the VM's local
IP address, the gateway address, MAC addresses for both the virtual Ethernet
device and physical Ethernet device, ARP tables, etc. All of these must be
valid for each function instance restored from a snapshot to avoid requiring
DHCP to discover new IP addresses, ARP to discover routes, and other dynamic
network configuration tasks.

To accomplish this, the network initialization code in the \rootfs configures
the network in accordance with strong guarantees from the hypervisor regarding the
network. VMs use a static local IP address connected to a virtual bridge device
on the hypervisor with a fixed IP address and MAC address. And this state is
captured in the \base snapshot which is shared across different function
instances. In order for different VMs on the same host to use the same local IP
addresses, VMs' virtual network devices are attached to a software Ethernet
bridge on the host.

This design allows 1) that any VM to use the bridge as
the gateway device which has a well-known IP address and 2) that different
VMs on the same host with the same IP address to communicate with the outside
since an Ethernet bridge uses hardware MAC addresses as identification.~\footnote{Note
that this setup only enables VMs to initiate connections, which is desirable
as FaaS functions are not network addressed like conventional servers.}

\section{Implementation}\label{sec:impl}

\begin{figure}[t]
    \centering
    \includegraphics[width=\linewidth]{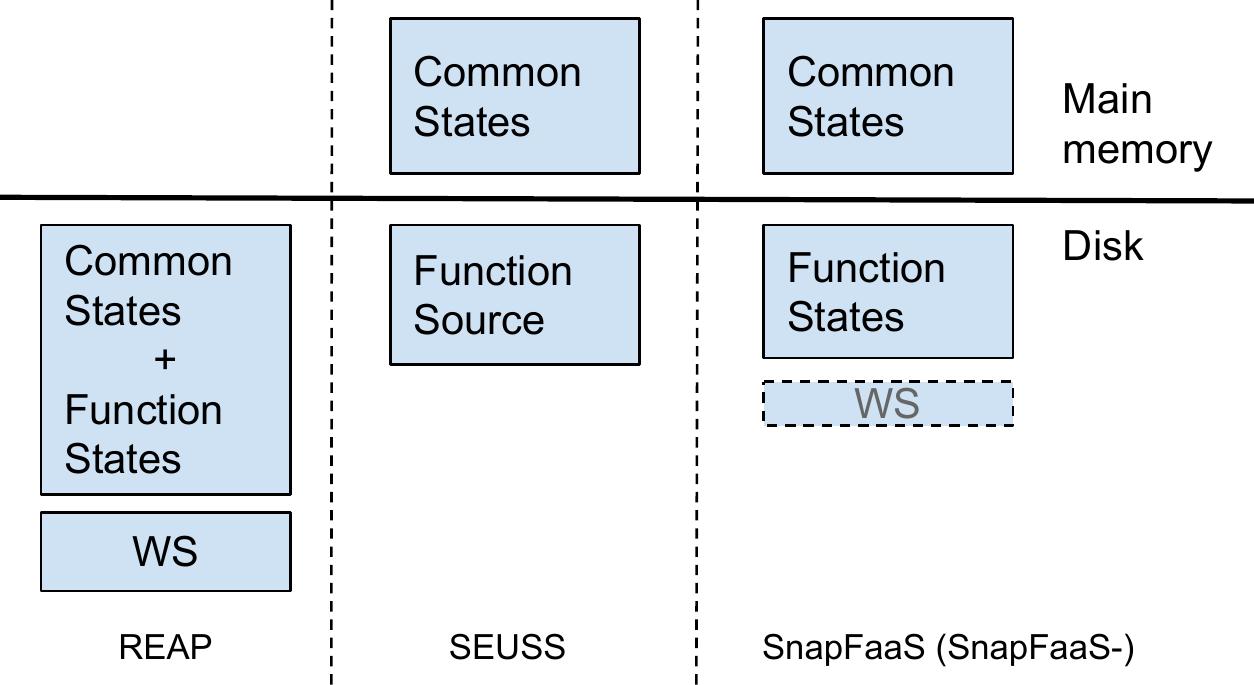}
    \caption{Comparison of \name{} with the existing snapshots for cold-starts.
    REAP captures the execution state as a whole and eagerly loads
    the working set.
    SEUSS restores cached common states on demand
    and then import functions from the source.
    \name{} caches common states and store function states on disk and eagerly loads
    only the working set. (\name{}- is \name{} without working set estimation.)}
    \label{fig:snapshot-overview}
\end{figure}

\begin{figure}[t]
  \begin{subfigure}[c]{\columnwidth}
    \begin{minted}[gobble=6]{python}
      import lorem
      def handle(event):
        return {
          'body': lorem.sentence(),
        }
    \end{minted}
      \caption{Entry-point procedure \texttt{handle}.
      The runtime starts function execution by
      calling \texttt{handle}. \texttt{handle} requires one parameter which
      contains any runtime arguments and returns an object.}
      \label{fig:source-code}
  \end{subfigure}
  \hfill
  \begin{subfigure}[c]{\columnwidth}
    \dirtree{%
      .1 /.
      .2 workload.
      .2 package/.
      .3 lorem/.
      .4 \_\_init\_\_.py.
      .4 data.py.
      .4 text.py.
      .2 lib/.
      .2 bin/.
    }
    \caption{Tarball structure.
      \texttt{workload} should contain procedure \texttt{handle}.
      Packages, libraries, and binaries should
      be in directory \texttt{package}, \texttt{lib}, and \texttt{bin}, respectively.
      }
    \label{fig:tarball-structure}
    \end{subfigure}
  \caption{\name{} programming interface (Python3 \texttt{lorem} example).
  This figure shows the content of the tarball that the developer submits to \name{} to register a function.}
  \label{fig:example-app-structure}
\end{figure}

\begin{figure*}[t!]
  \centering
  \includegraphics[width=\textwidth]{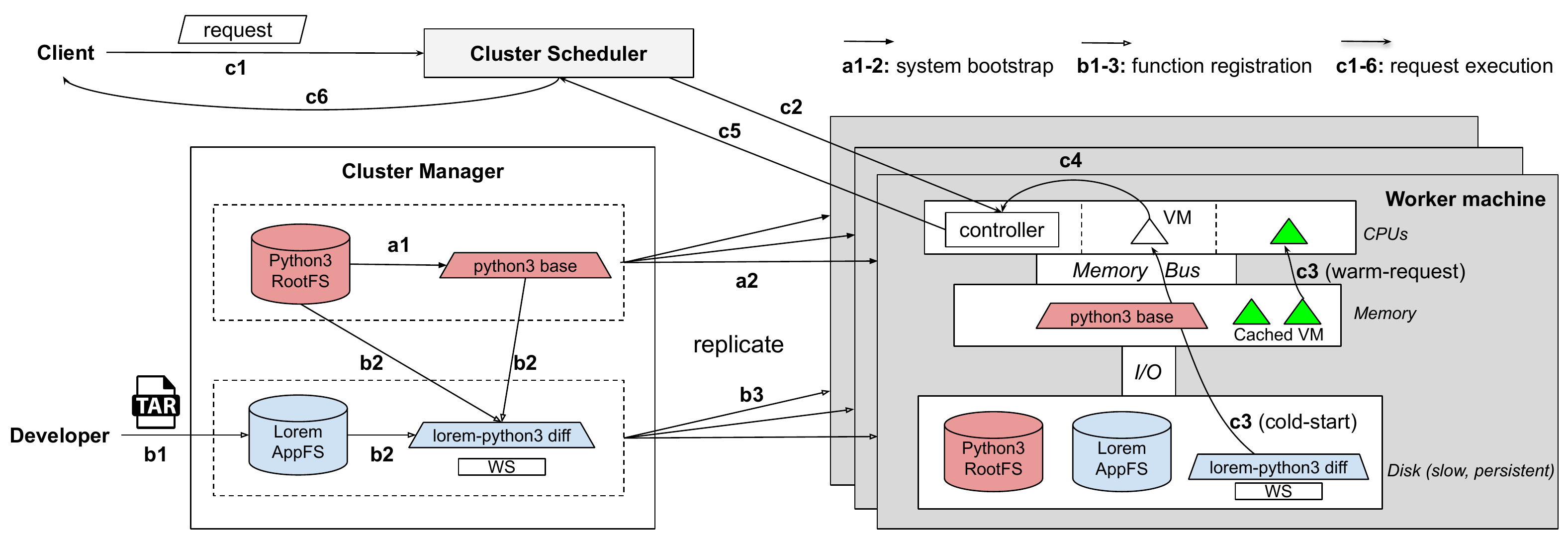}
  \caption{Overview of a FaaS system with \name{} deployed. 
  }
  \label{fig:architecture}
\end{figure*}
\subsection{Software Stack}\label{sec:impl:softwarestack}
In addition to Firecracker microVMs in which user code runs, there are other components
along the software stack. Precisely, each VM runs alongside its Firecracker virtual machine manager (VMM) process
and on top of Linux KVM hypervisor~\cite{kvm}. Linux KVM virtualizes CPU and memory
and the VMM process assists with I/Os.

\paragraph{VM Pre-configuration.}
The VMM process is also responsible for pre-configuring a VM.
Before booting a VM from either the kernel or snapshots, the system needs
to pre-configure it, including registering a new VM with Linux KVM asking for CPUs \&
memory and attaching virtual I/O devices to the VM after registration.

\paragraph{VM Organization.}
In addition to a stripped-down Linux kernel, two block devices formatted
to \rootfs and \appfs and an IP network device, each VM has a
VSOCK~\cite{virtio} (virtual socket) device to communicate with the host, \ie
receiving runtime arguments and returning results.

We follow the guide provided by the Firecracker team about how to build file
system images~\cite{firecracker-github}.
File system images are based on Alpine Linux 3.10~\cite{alpine}---a distribution
of Linux that uses the lightweight BusyBox UNIX utilities and OpenRC~\cite{openrc}
\texttt{init} system---and a Linux kernel based on version 4.20 compiled with the minimal
configurations.

\subsection{\name{} Generation and Restoration}\label{sec:impl:snapshot}
\paragraph{Generation.}
The VMM process and the custom language-specific \emph{runtime entry point} running
inside the VM cooperatively capture the VM's execution state. A custom language-specific
\emph{runtime entry point} is a script (for scripting languages like Python and JavaScript)
or an executable (for compiled languages like Go) that when executed initializes
the language runtime including executing basic library code used by itself.
It also mounts the \appfs.
More importantly, it makes hypercalls that pause the VM and cause a context switch
from the VM in guest mode to the VMM in host user mode.

To create a \base snapshot, we boot a VM normally from the kernel using the \rootfs,
with a placeholder \appfs image. The VMM enables dirty-page tracking in order to tell
which pages must be recorded in the snapshot. Once language runtime initialization
completes, the language-specific runtime entry point immediately makes a hypercall
from each virtual CPU core. This action pause the VM and signals the VMM to capture
the VM's execution state, \ie the state of memory, CPU, and I/O devices and generate
the \base snapshot. This snapshot consists of a sparse file containing only dirty
memory pages, as well as a JSON file describing non-memory state, the state of CPU and
I/O devices. The resulting snapshot is stored in a \texttt{tmpfs} file system on the host
for easy in-memory storage.

To creates a \diff snapshot for a particular function, we begin by booting a VM
through restoring the correct \base snapshot with the function's actual \appfs
attached to the VM instead. As a result, the VM continues
execution inside the language-specific runtime entry point immediately after the
hypercall described above. The runtime entry point continues by mounting the
\appfs and loading and initializing the function in a language-specific
way---\eg in Python and Node.js, importing the \texttt{workload} file is
sufficient to execute initialization code in the global scope (such as
importing third-party libraries) while the Go entry point loads an ELF
``plugin'' and explicitly calls an \texttt{Init} that the function is expected
to export.

Once the function is initialized, without actually invoking it, the runtime
entry point pauses the VM and signals the VMM to generate the \diff snapshot by,
again, making a hypercall from each virtual CPU core. During this process, the VMM
enables dirty page tracking as well and \diff snapshots contain only pages that are
marked dirty during this time. For memory state, in addition to dirty memory pages themselves,
a \diff snapshot includes an explicit record of dirty pages relative to the \base snapshot.
Restoration from \base and \diff snapshots uses this metadata. For non-memory state,
\diff snapshots use a similar JSON file.

To generate a WS file, similar to generate a \diff snapshot, we begin by
restoring \base and \diff snapshots all on-demand to allow us to track page
access. After execution, we record the set of pages that are accessed during execution
and are \texttt{file-mmap}ed from the \diff snapshot. This set is the working set of
the \diff snapshot.

\paragraph{Restoration.}
To restore the VM's memory, the VMM \emph{file-mmap}s the \base snapshot into
private memory. As a result, memory pages from the \base snapshot are loaded
copy-on-write, and any pages not modified by function execution will be shared
among VMs running the same language runtime. The VMM then \emph{copies} each
page in the \diff snapshot into private memory using the system call \texttt{readv}.

Or if the working set optimization is applied, the VMM \emph{file-mmap}s
non-working-set pages of the function's \diff snapshot into private memory
and \emph{copies} the working set into the VM's private memory.

To restore the non-memory states, the VMM restores non-memory states encoded in the
\diff snapshot's JSON file, including the state of CPU and I/O devices.

\subsection{\name{} Deployed}
\paragraph{Programming Interface.}
\name{} assumes a similar programming interface to existing public FaaS
offerings~\cite{aws-lambda}. To register a function, the
developer submits a tarball of the function source including all dependencies.
The system then reformats the tarball to a \appfs.

Figure~\ref{fig:example-app-structure} is an example of the function source
and the tarball structure. Developers should define a procedure
named \texttt{handle} which is the entry point of function execution. 
\texttt{handle} takes one argument \texttt{event}
and returns a serializable object. The \texttt{event} argument contains inputs if any.
Figure~\ref{fig:source-code} shows an example of a simple Python3 function.

To prepare a tarball, developers should name the file that defines \texttt{handle}
as \texttt{workload} (Figure~\ref{fig:tarball-structure}).
Additionally, third-party dependencies, native libraries and native binaries are in
subdirectories \texttt{package}, \texttt{lib}, and \texttt{bin}, respectively.

\paragraph{System Workflow.}
Figure~\ref{fig:architecture} shows how a FaaS system employing SnapFaaS looks like:
a cluster scheduler, a cluster manager\footnote{The cluster scheduler and the cluster manager are not within
this paper's scope.}, and a fleet of worker machines each
of which has two-tiered storage: memory and slow, persistent disk. Memory
stores \base snapshots while disk stores WS files, \diff snapshots and file system images.

System workflow is as follows.

During system bootstrap, for each supported language, 
the cluster manager generates a \rootfs image for each supported language that
serves as the boot device. Next, the cluster manager generates a \base snapshot
using the language's \rootfs image. Finally, the cluster manager
replicates these \rootfs-es and \base snapshots to each worker machine's disk
and memory respectively.

At function registrations and updates, the cluster manager converts tarballs into
\appfs images and generates \diff snapshots
using \appfs images with the correct \base snapshot and
\rootfs image. Then the manager invokes functions with mock arguments to create
WS files. At last, the cluster manager
replicates these WS files, \diff snapshots and \appfs images to each worker
machine's disk. \appfs images need replicating because the function may
dynamically loads packages or invokes binaries that's stored on the \appfs.

For client requests, the cluster scheduler is the gateway. Upon receiving a request,
the cluster scheduler dispatches it to a worker machine.
The controller running on the worker machine either invokes the
requested function in a cached idle function instance (warm-request) or launch
a new instance (cold-start) in a VM with the correct \base and \diff snapshots plus
WS files.
Once the function finishes execution, it sends the results to the controller.
The controller then returns the response to the client.
\section{Evaluation}\label{sec:eval}
\begin{table*}[h]
    \centering
    \resizebox{\textwidth}{!}{%
    \begin{tabular}{llll}
    \hline
    \textbf{Name} & \textbf{Description} & \textbf{Language} & \textbf{Libraries \& Binaries} \\ \hline
    lorem & Generate a random lorem text string & Node.js, Python3, Go & lorem \\
    sentiment-analysis & Textual sentiment analysis with NLP models & Python3, Go & nltk, textblob etc.\\
    thumbnail & Generate a thumbnail picture & Python3, Java & PIL/ImageMagick, libjpeg etc. \\
    ocr & Text recognition with Tesseract OCR & Node.js & tesseract, tessdata, libjpeg etc. \\
    img-resize & Resize a large image to 5 smaller sizes & Node.js & jimp, node-zip \\
    alexa-door & Control door lock with Alexa & Node.js & ask-sdk-core, request etc. \\
    alexa-reminder & Setting up reminders with Alexa & Node.js & ask-sdk-core, request etc. \\
    audio-fingerprint & Generate acoustic fingerprints of audio files & Python3 & pyacoustid, audioread etc.\\
    matmul & Matrix multiplication & Java & None\\
    tpcc & TPC-C benchmark & Java & java.sql \\ \hline
    \end{tabular}%
    }
    \caption[]{Benchmarking functions}
    \label{tab:applications}
\end{table*}

We evaluate \name{} by answering the following questions:

\begin{itemize}[noitemsep]
    \item How do \snapfaas and \snapfaasnoprof (\name{} without working set approximation)
    perform compared with the existing snapshots?

    \item What are cold-start overhead breakdowns like for different snapshots
    following the model we proposed in Section~\ref{sec:model}?
\end{itemize}

\subsection{Experimental Setup}\label{sec:eval-setup}

\paragraph{Hardware.}
We use CloudLab's~\cite{cloudlab} c220g5 machines. A c220g5 has two Intel Xeon
10-core CPUs at 2.20 GHz, DDR4-2666 memory, one SATA SSD with 500 MB/s peak
sequential read bandwidth and 50 us random read latency. We disable
hyperthreading~\cite{firecracker-nsdi}. The host operating system is Ubuntu
16.04.1 with kernel version 4.15.0 and the guest operating system is Alpine
Linux 3.10 with kernel version 4.20.0.

\paragraph{Benchmarking functions.}
We implemented 14 functions in four languages\footnote{All applications are
available at \url{URL removed for anonymity}.} (Table~\ref{tab:applications})
that represent a variety of common FaaS applications: text,
audio, and image processing; online transaction processing; and smarthome/IoT
applications. 

Many of these applications have library dependencies, including native libraries
and executables. For example, the Python3 \texttt{thumbnail} function depends
on the \texttt{Pillow} package which requires the \texttt{libjpeg} native
library. The Node.js \texttt{ocr} function is a thin wrapper around
\texttt{Tesseract OCR} executable, and the Node.js \texttt{alexa-reminder}
function adds and retrieves reminder items stored in CouchDB over the network.

\paragraph{\seuss and \reap}
In order to compare \name{} with the existing snapshots,
we implemented \name{} versions of them, \reap and \seuss.
We refer to \name{} version's REAP~\cite{reap} as \reap and implement it as
eagerly loading the working set and demand-paging the rest of a full-function
snapshot. We refer to \name{} version's SEUSS-like designs~\cite{seuss, catalyzer, sock} 
as \seuss and implement it as copy-on-write sharing a in-memory base snapshot
and importing the function from its on-disk source.


\paragraph{Function inputs.} For \snapfaas and \reap experiments, we use the
same function inputs that we use to generate the working sets.


\begin{figure*}[t!]
    \centering
    \begin{subfigure}[t]{0.49\textwidth}
        \centering
        \includegraphics[width=\linewidth]{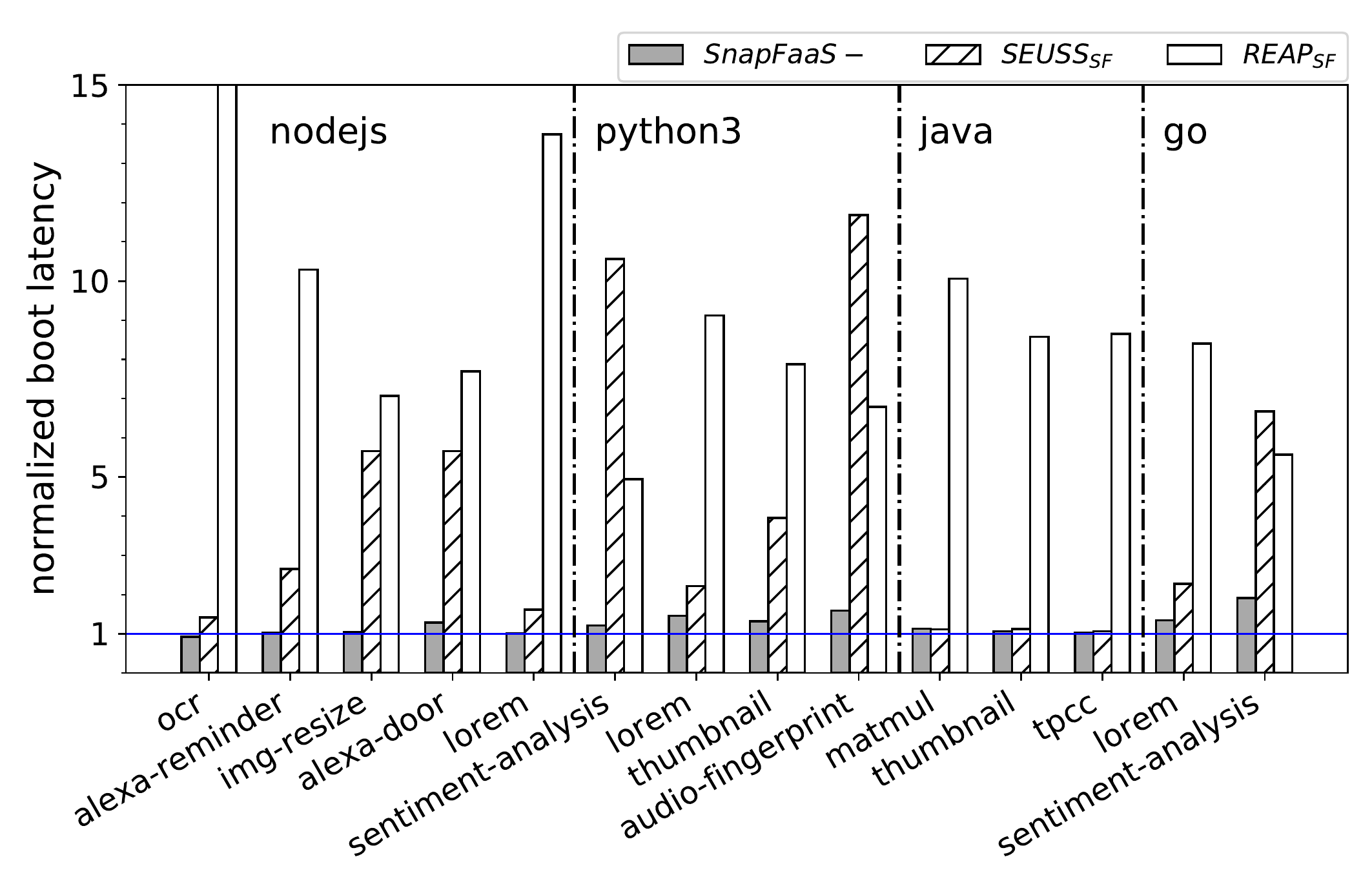}
        \caption{Boot latency normalized to \snapfaas.}
        \label{fig:microbench-boot}
    \end{subfigure}
    \begin{subfigure}[t]{0.49\textwidth}
        \centering
        \includegraphics[width=\linewidth]{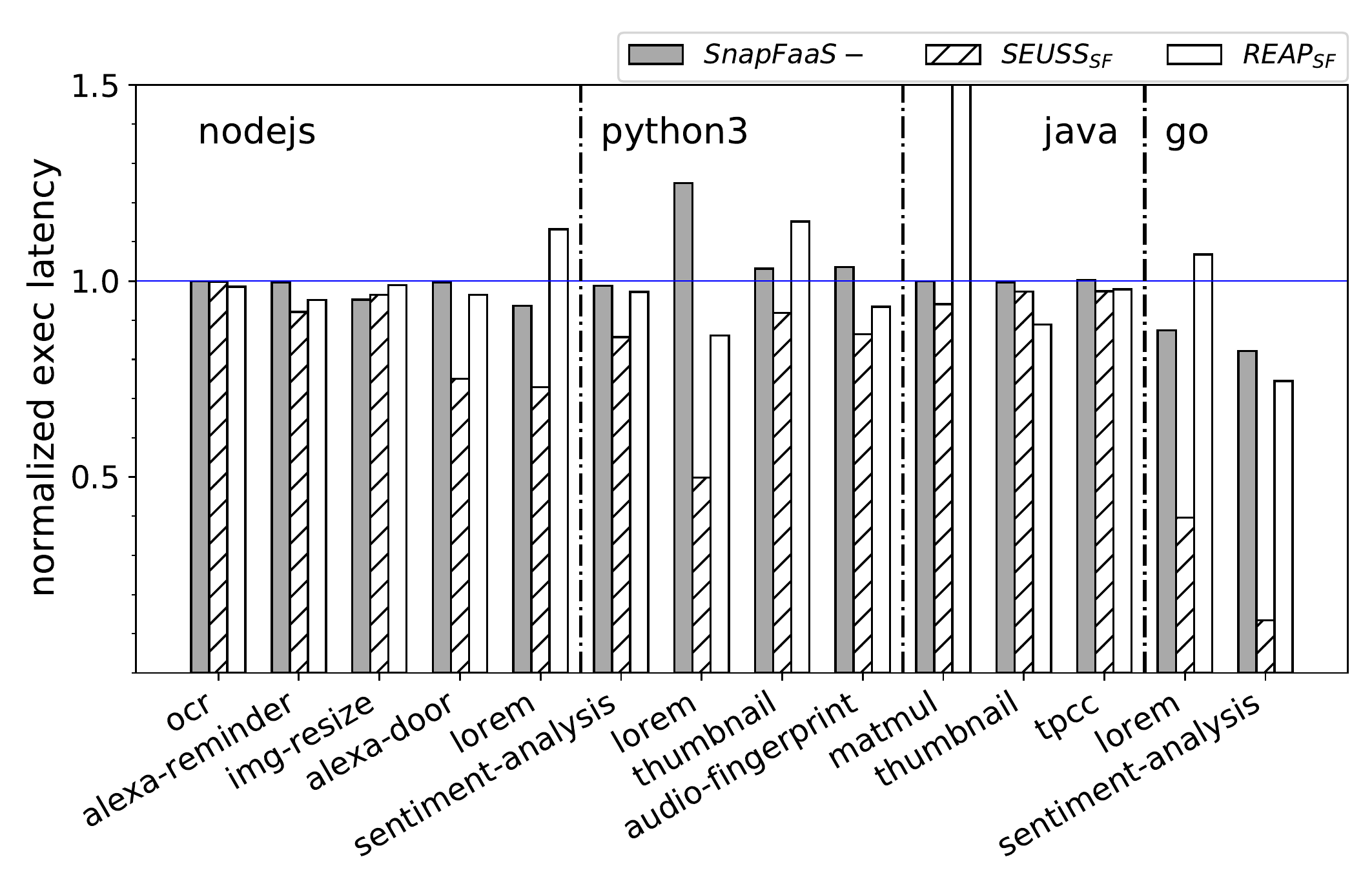}
        \caption{Execution latency normalized to \snapfaas.}
        \label{fig:microbench-exec}
    \end{subfigure}
    \begin{subfigure}[t]{0.49\textwidth}
        \centering
        \includegraphics[width=\linewidth]{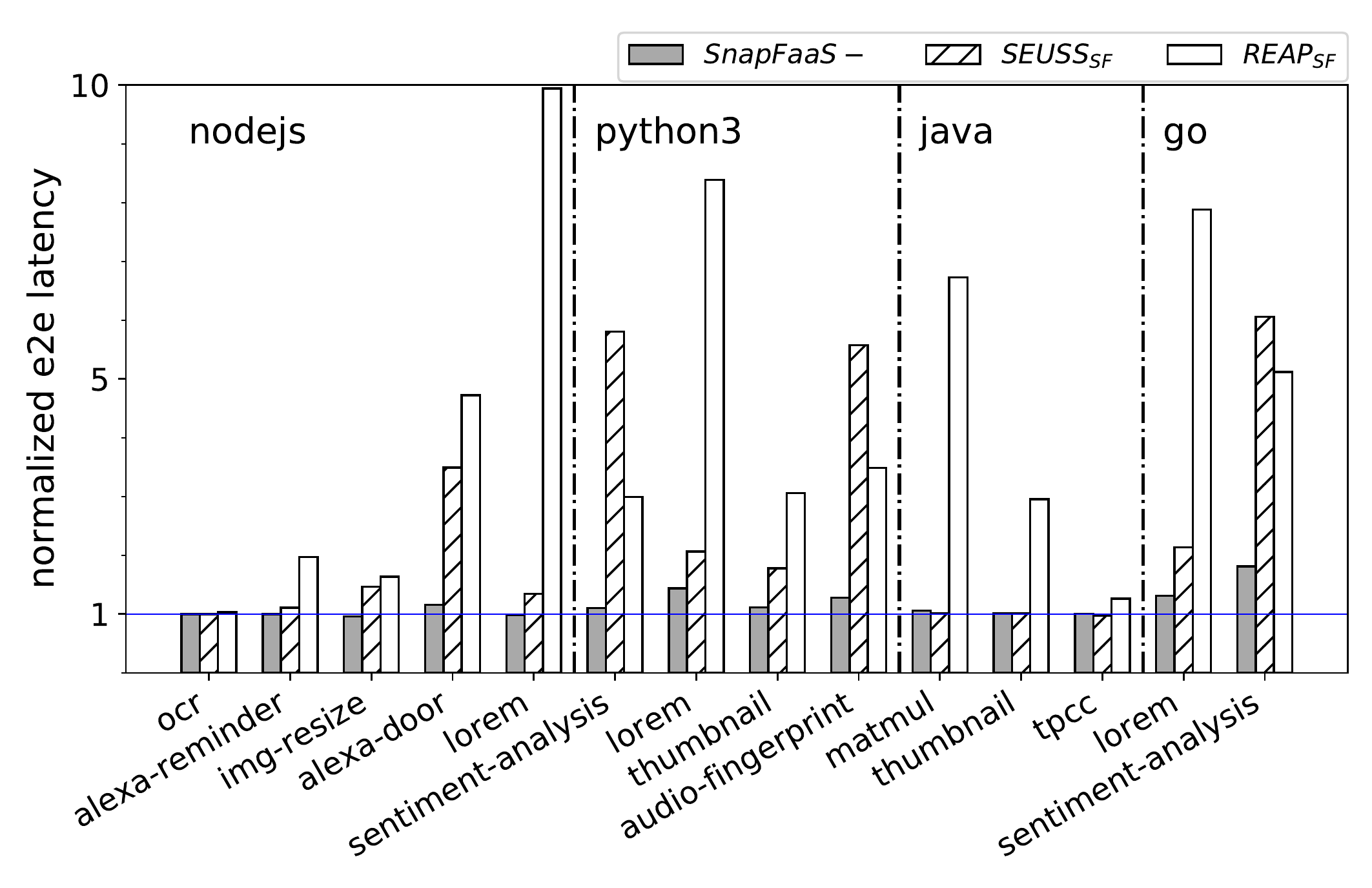}
        \caption{End-to-end latency normalized to \snapfaas.}
        \label{fig:microbench-e2e}
    \end{subfigure}
    \begin{subfigure}[t]{0.49\textwidth}
        \centering
        \includegraphics[width=\linewidth]{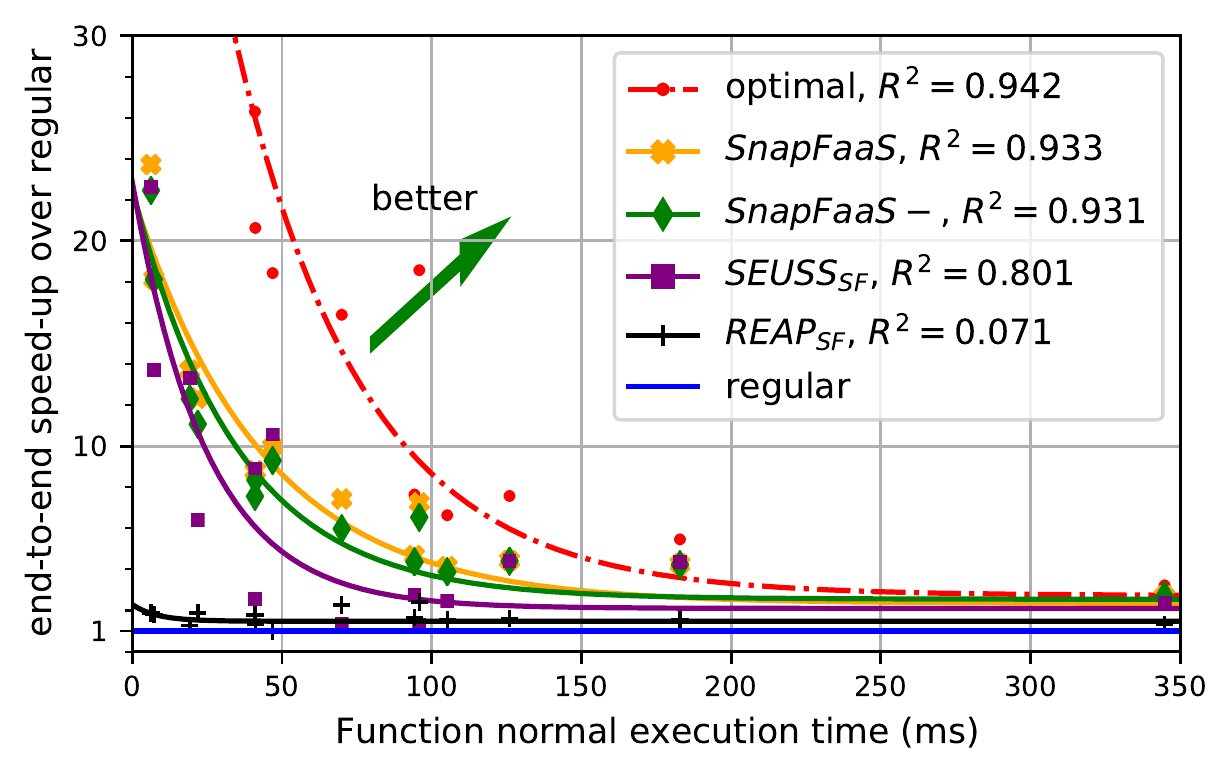}
        \caption{\snapfaas, \snapfaasnoprof, \seuss, and \reap end-to-end \textbf{speed-up}
        over \regular vs \regular's function execution time. \regular stands for
        booting a function through the regular VM booting process. \texttt{optimal} includes
        \emph{only} the execution time of warm functions and thus represents
        the speed-up of an optimal cold-start strategy.}
        \label{fig:speed-up-trend}
    \end{subfigure}
    \caption{Snapshot performance comparison. We take the latency average
    for 100 rounds and normalize \snapfaas's latencies to 1 (the blue line).
    For figure (a) - (c), being above the blue line means being slower than \snapfaas.}
    \label{fig:microbench}
\end{figure*}

\subsection{Snapshot Performance Comparison}\label{sec:eval-performance}

Figure~\ref{fig:microbench} compares the cold-start latency of \snapfaasnoprof and
\snapfaas with \reap and \seuss.

\paragraph{Cold-Start Boot Latency.}\label{sec:eval-boot}

We measure boot time for cold-start requests from when the VMM process starts
to when the VM is ready to accept client requests.
Figure~\ref{fig:microbench-boot} shows boot latencies normalized to
\snapfaas.

Without the working set optimization, it is already the case that \snapfaasnoprof is always
faster than \reap and always at least comparable to \seuss.
Specifically, \snapfaasnoprof is up to 20.2x as fast as \reap and up to 8.8x as fast as \seuss.

With the working set optimization, \snapfaas is always at least comparable to \snapfaasnoprof, \reap and \seuss.
Specifically, \snapfaas is up to 1.9x as fast as \snapfaasnoprof, up to 18.8x as fast as \reap, and
up to 12x as fast as \seuss.

\begin{table*}[]
    \centering
    \footnotesize
    \begin{tabular}{lrrrrrrrrrrrrrrrr}
    \multicolumn{1}{c}{}        & \multicolumn{4}{c}{\reap}                                                                                   & \multicolumn{4}{c}{\seuss}                                                                                  & \multicolumn{4}{c}{\snapfaasnoprof}                                                                               & \multicolumn{4}{c}{\snapfaas}                                                                             \\
                                & \multicolumn{1}{l}{A}                        & \multicolumn{1}{l}{B}                        & \multicolumn{1}{l}{C}                        & \multicolumn{1}{l}{D}                        & \multicolumn{1}{l}{A}                        & \multicolumn{1}{l}{B}                        & \multicolumn{1}{l}{C}                        & \multicolumn{1}{l}{D}                        & \multicolumn{1}{l}{A}                        & \multicolumn{1}{l}{B}                        & \multicolumn{1}{l}{C}                        & \multicolumn{1}{l}{D}                        & \multicolumn{1}{l}{A}                        & \multicolumn{1}{l}{B}                        & \multicolumn{1}{l}{C}                        & \multicolumn{1}{l}{D}                        \\ \hline
    \multicolumn{1}{r}{Go}      & \multicolumn{1}{l}{\cellcolor[HTML]{9B9B9B}} & \multicolumn{1}{l}{\cellcolor[HTML]{C0C0C0}} & \multicolumn{1}{l}{\cellcolor[HTML]{EFEFEF}} & \multicolumn{1}{l}{\cellcolor[HTML]{FFFFFF}} & \multicolumn{1}{l}{\cellcolor[HTML]{9B9B9B}} & \multicolumn{1}{l}{\cellcolor[HTML]{C0C0C0}} & \multicolumn{1}{l}{\cellcolor[HTML]{EFEFEF}} & \multicolumn{1}{l}{\cellcolor[HTML]{FFFFFF}} & \multicolumn{1}{l}{\cellcolor[HTML]{9B9B9B}} & \multicolumn{1}{l}{\cellcolor[HTML]{C0C0C0}} & \multicolumn{1}{l}{\cellcolor[HTML]{EFEFEF}} & \multicolumn{1}{l}{\cellcolor[HTML]{FFFFFF}} & \multicolumn{1}{l}{\cellcolor[HTML]{9B9B9B}} & \multicolumn{1}{l}{\cellcolor[HTML]{C0C0C0}} & \multicolumn{1}{l}{\cellcolor[HTML]{EFEFEF}} & \multicolumn{1}{l}{\cellcolor[HTML]{FFFFFF}} \\
    lorem                       & \cellcolor[HTML]{9B9B9B}6.9                  & \cellcolor[HTML]{C0C0C0}160.3                & \cellcolor[HTML]{EFEFEF}9.1                  & \cellcolor[HTML]{FFFFFF}0.9                  & \cellcolor[HTML]{9B9B9B}4.6                  & \cellcolor[HTML]{C0C0C0}0.6                  & \cellcolor[HTML]{EFEFEF}42.4                 & \cellcolor[HTML]{FFFFFF}-0.2                 & \cellcolor[HTML]{9B9B9B}5.5                  & \cellcolor[HTML]{C0C0C0}19.7                 & \cellcolor[HTML]{EFEFEF}3.1                  & \cellcolor[HTML]{FFFFFF}0.6                  & \cellcolor[HTML]{9B9B9B}5.7                  & \cellcolor[HTML]{C0C0C0}10.3                 & \cellcolor[HTML]{EFEFEF}5.0                  & \cellcolor[HTML]{FFFFFF}0.8                  \\
    sentiment-analysis          & \cellcolor[HTML]{9B9B9B}7.7                  & \cellcolor[HTML]{C0C0C0}448.5                & \cellcolor[HTML]{EFEFEF}4.8                  & \cellcolor[HTML]{FFFFFF}5.4                  & \cellcolor[HTML]{9B9B9B}4.6                  & \cellcolor[HTML]{C0C0C0}0.7                  & \cellcolor[HTML]{EFEFEF}546.7                & \cellcolor[HTML]{FFFFFF}0.2                  & \cellcolor[HTML]{9B9B9B}7.3                  & \cellcolor[HTML]{C0C0C0}144.9                & \cellcolor[HTML]{EFEFEF}6.0                  & \cellcolor[HTML]{FFFFFF}6.0                  & \cellcolor[HTML]{9B9B9B}6.9                  & \cellcolor[HTML]{C0C0C0}69.9                 & \cellcolor[HTML]{EFEFEF}5.9                  & \cellcolor[HTML]{FFFFFF}7.6                  \\ \hline
    \multicolumn{1}{r}{Java}    & \multicolumn{1}{l}{\cellcolor[HTML]{9B9B9B}} & \multicolumn{1}{l}{\cellcolor[HTML]{C0C0C0}} & \multicolumn{1}{l}{\cellcolor[HTML]{EFEFEF}} & \multicolumn{1}{l}{\cellcolor[HTML]{FFFFFF}} & \multicolumn{1}{l}{\cellcolor[HTML]{9B9B9B}} & \multicolumn{1}{l}{\cellcolor[HTML]{C0C0C0}} & \multicolumn{1}{l}{\cellcolor[HTML]{EFEFEF}} & \multicolumn{1}{l}{\cellcolor[HTML]{FFFFFF}} & \multicolumn{1}{l}{\cellcolor[HTML]{9B9B9B}} & \multicolumn{1}{l}{\cellcolor[HTML]{C0C0C0}} & \multicolumn{1}{l}{\cellcolor[HTML]{EFEFEF}} & \multicolumn{1}{l}{\cellcolor[HTML]{FFFFFF}} & \multicolumn{1}{l}{\cellcolor[HTML]{9B9B9B}} & \multicolumn{1}{l}{\cellcolor[HTML]{C0C0C0}} & \multicolumn{1}{l}{\cellcolor[HTML]{EFEFEF}} & \multicolumn{1}{l}{\cellcolor[HTML]{FFFFFF}} \\
    matmul                      & \cellcolor[HTML]{9B9B9B}7.1                  & \cellcolor[HTML]{C0C0C0}417.2                & \cellcolor[HTML]{EFEFEF}17.7                 & \cellcolor[HTML]{FFFFFF}186.5                & \cellcolor[HTML]{9B9B9B}5.0                  & \cellcolor[HTML]{C0C0C0}1.0                  & \cellcolor[HTML]{EFEFEF}43.0                 & \cellcolor[HTML]{FFFFFF}11.2                 & \cellcolor[HTML]{9B9B9B}6.3                  & \cellcolor[HTML]{C0C0C0}28.0                 & \cellcolor[HTML]{EFEFEF}15.4                 & \cellcolor[HTML]{FFFFFF}14.4                 & \cellcolor[HTML]{9B9B9B}6.3                  & \cellcolor[HTML]{C0C0C0}23.9                 & \cellcolor[HTML]{EFEFEF}13.7                 & \cellcolor[HTML]{FFFFFF}14.4                 \\
    thumbnail                   & \cellcolor[HTML]{9B9B9B}7.5                  & \cellcolor[HTML]{C0C0C0}436.3                & \cellcolor[HTML]{EFEFEF}24.9                 & \cellcolor[HTML]{FFFFFF}6.2                  & \cellcolor[HTML]{9B9B9B}4.5                  & \cellcolor[HTML]{C0C0C0}0.7                  & \cellcolor[HTML]{EFEFEF}55.9                 & \cellcolor[HTML]{FFFFFF}18.8                 & \cellcolor[HTML]{9B9B9B}5.9                  & \cellcolor[HTML]{C0C0C0}29.4                 & \cellcolor[HTML]{EFEFEF}22.6                 & \cellcolor[HTML]{FFFFFF}22.2                 & \cellcolor[HTML]{9B9B9B}6.1                  & \cellcolor[HTML]{C0C0C0}26.6                 & \cellcolor[HTML]{EFEFEF}22.0                 & \cellcolor[HTML]{FFFFFF}22.7                 \\
    tpcc                        & \cellcolor[HTML]{9B9B9B}6.9                  & \cellcolor[HTML]{C0C0C0}408.3                & \cellcolor[HTML]{EFEFEF}16.4                 & \cellcolor[HTML]{FFFFFF}102.7                & \cellcolor[HTML]{9B9B9B}5.1                  & \cellcolor[HTML]{C0C0C0}0.8                  & \cellcolor[HTML]{EFEFEF}47.1                 & \cellcolor[HTML]{FFFFFF}96.3                 & \cellcolor[HTML]{9B9B9B}6.5                  & \cellcolor[HTML]{C0C0C0}29.7                 & \cellcolor[HTML]{EFEFEF}14.7                 & \cellcolor[HTML]{FFFFFF}134.0                & \cellcolor[HTML]{9B9B9B}6.5                  & \cellcolor[HTML]{C0C0C0}28.7                 & \cellcolor[HTML]{EFEFEF}14.7                 & \cellcolor[HTML]{FFFFFF}130.1                \\ \hline
    \multicolumn{1}{r}{Node.js} & \multicolumn{1}{l}{\cellcolor[HTML]{9B9B9B}} & \multicolumn{1}{l}{\cellcolor[HTML]{C0C0C0}} & \multicolumn{1}{l}{\cellcolor[HTML]{EFEFEF}} & \multicolumn{1}{l}{\cellcolor[HTML]{FFFFFF}} & \multicolumn{1}{l}{\cellcolor[HTML]{9B9B9B}} & \multicolumn{1}{l}{\cellcolor[HTML]{C0C0C0}} & \multicolumn{1}{l}{\cellcolor[HTML]{EFEFEF}} & \multicolumn{1}{l}{\cellcolor[HTML]{FFFFFF}} & \multicolumn{1}{l}{\cellcolor[HTML]{9B9B9B}} & \multicolumn{1}{l}{\cellcolor[HTML]{C0C0C0}} & \multicolumn{1}{l}{\cellcolor[HTML]{EFEFEF}} & \multicolumn{1}{l}{\cellcolor[HTML]{FFFFFF}} & \multicolumn{1}{l}{\cellcolor[HTML]{9B9B9B}} & \multicolumn{1}{l}{\cellcolor[HTML]{C0C0C0}} & \multicolumn{1}{l}{\cellcolor[HTML]{EFEFEF}} & \multicolumn{1}{l}{\cellcolor[HTML]{FFFFFF}} \\
    alexa-door                  & \cellcolor[HTML]{9B9B9B}9.0                  & \cellcolor[HTML]{C0C0C0}528.8                & \cellcolor[HTML]{EFEFEF}14.3                 & \cellcolor[HTML]{FFFFFF}13.7                 & \cellcolor[HTML]{9B9B9B}7.6                  & \cellcolor[HTML]{C0C0C0}1.1                  & \cellcolor[HTML]{EFEFEF}397.3                & \cellcolor[HTML]{FFFFFF}1.6                  & \cellcolor[HTML]{9B9B9B}9.0                  & \cellcolor[HTML]{C0C0C0}70.4                 & \cellcolor[HTML]{EFEFEF}12.6                 & \cellcolor[HTML]{FFFFFF}15.5                 & \cellcolor[HTML]{9B9B9B}8.7                  & \cellcolor[HTML]{C0C0C0}50.9                 & \cellcolor[HTML]{EFEFEF}12.2                 & \cellcolor[HTML]{FFFFFF}15.7                 \\
    alexa-reminder              & \cellcolor[HTML]{9B9B9B}6.9                  & \cellcolor[HTML]{C0C0C0}444.4                & \cellcolor[HTML]{EFEFEF}15.2                 & \cellcolor[HTML]{FFFFFF}7.9                  & \cellcolor[HTML]{9B9B9B}4.6                  & \cellcolor[HTML]{C0C0C0}0.7                  & \cellcolor[HTML]{EFEFEF}114.7                & \cellcolor[HTML]{FFFFFF}-3.2                 & \cellcolor[HTML]{9B9B9B}6.3                  & \cellcolor[HTML]{C0C0C0}33.0                 & \cellcolor[HTML]{EFEFEF}7.0                  & \cellcolor[HTML]{FFFFFF}24.6                 & \cellcolor[HTML]{9B9B9B}6.2                  & \cellcolor[HTML]{C0C0C0}31.9                 & \cellcolor[HTML]{EFEFEF}7.2                  & \cellcolor[HTML]{FFFFFF}25.9                 \\
    img-resize                  & \cellcolor[HTML]{9B9B9B}7.6                  & \cellcolor[HTML]{C0C0C0}507.0                & \cellcolor[HTML]{EFEFEF}14.7                 & \cellcolor[HTML]{FFFFFF}33.7                 & \cellcolor[HTML]{9B9B9B}4.7                  & \cellcolor[HTML]{C0C0C0}1.0                  & \cellcolor[HTML]{EFEFEF}417.9                & \cellcolor[HTML]{FFFFFF}18.2                 & \cellcolor[HTML]{9B9B9B}6.4                  & \cellcolor[HTML]{C0C0C0}61.1                 & \cellcolor[HTML]{EFEFEF}10.4                 & \cellcolor[HTML]{FFFFFF}10.4                 & \cellcolor[HTML]{9B9B9B}7.2                  & \cellcolor[HTML]{C0C0C0}57.9                 & \cellcolor[HTML]{EFEFEF}9.8                  & \cellcolor[HTML]{FFFFFF}40.2                 \\
    lorem                       & \cellcolor[HTML]{9B9B9B}7.5                  & \cellcolor[HTML]{C0C0C0}374.5                & \cellcolor[HTML]{EFEFEF}19.1                 & \cellcolor[HTML]{FFFFFF}7.0                  & \cellcolor[HTML]{9B9B9B}4.7                  & \cellcolor[HTML]{C0C0C0}0.7                  & \cellcolor[HTML]{EFEFEF}41.7                 & \cellcolor[HTML]{FFFFFF}1.9                  & \cellcolor[HTML]{9B9B9B}7.6                  & \cellcolor[HTML]{C0C0C0}16.8                 & \cellcolor[HTML]{EFEFEF}5.1                  & \cellcolor[HTML]{FFFFFF}4.5                  & \cellcolor[HTML]{9B9B9B}7.6                  & \cellcolor[HTML]{C0C0C0}14.4                 & \cellcolor[HTML]{EFEFEF}7.2                  & \cellcolor[HTML]{FFFFFF}5.3                  \\
    ocr                         & \cellcolor[HTML]{9B9B9B}7.1                  & \cellcolor[HTML]{C0C0C0}466.0                & \cellcolor[HTML]{EFEFEF}14.6                 & \cellcolor[HTML]{FFFFFF}133.7                & \cellcolor[HTML]{9B9B9B}4.5                  & \cellcolor[HTML]{C0C0C0}0.7                  & \cellcolor[HTML]{EFEFEF}34.4                 & \cellcolor[HTML]{FFFFFF}255.9                & \cellcolor[HTML]{9B9B9B}5.0                  & \cellcolor[HTML]{C0C0C0}15.1                 & \cellcolor[HTML]{EFEFEF}5.6                  & \cellcolor[HTML]{FFFFFF}276.1                & \cellcolor[HTML]{9B9B9B}6.0                  & \cellcolor[HTML]{C0C0C0}15.0                 & \cellcolor[HTML]{EFEFEF}6.9                  & \cellcolor[HTML]{FFFFFF}280.2                \\ \hline
    \multicolumn{1}{r}{Python3} & \multicolumn{1}{l}{\cellcolor[HTML]{9B9B9B}} & \multicolumn{1}{l}{\cellcolor[HTML]{C0C0C0}} & \multicolumn{1}{l}{\cellcolor[HTML]{EFEFEF}} & \multicolumn{1}{l}{\cellcolor[HTML]{FFFFFF}} & \multicolumn{1}{l}{\cellcolor[HTML]{9B9B9B}} & \multicolumn{1}{l}{\cellcolor[HTML]{C0C0C0}} & \multicolumn{1}{l}{\cellcolor[HTML]{EFEFEF}} & \multicolumn{1}{l}{\cellcolor[HTML]{FFFFFF}} & \multicolumn{1}{l}{\cellcolor[HTML]{9B9B9B}} & \multicolumn{1}{l}{\cellcolor[HTML]{C0C0C0}} & \multicolumn{1}{l}{\cellcolor[HTML]{EFEFEF}} & \multicolumn{1}{l}{\cellcolor[HTML]{FFFFFF}} & \multicolumn{1}{l}{\cellcolor[HTML]{9B9B9B}} & \multicolumn{1}{l}{\cellcolor[HTML]{C0C0C0}} & \multicolumn{1}{l}{\cellcolor[HTML]{EFEFEF}} & \multicolumn{1}{l}{\cellcolor[HTML]{FFFFFF}} \\
    audio-fingerprint           & \cellcolor[HTML]{9B9B9B}7.3                  & \cellcolor[HTML]{C0C0C0}423.5                & \cellcolor[HTML]{EFEFEF}5.5                  & \cellcolor[HTML]{FFFFFF}7.7                  & \cellcolor[HTML]{9B9B9B}4.8                  & \cellcolor[HTML]{C0C0C0}0.6                  & \cellcolor[HTML]{EFEFEF}744.9                & \cellcolor[HTML]{FFFFFF}1.9                  & \cellcolor[HTML]{9B9B9B}7.4                  & \cellcolor[HTML]{C0C0C0}87.8                 & \cellcolor[HTML]{EFEFEF}6.9                  & \cellcolor[HTML]{FFFFFF}16.1                 & \cellcolor[HTML]{9B9B9B}7.4                  & \cellcolor[HTML]{C0C0C0}50.1                 & \cellcolor[HTML]{EFEFEF}6.8                  & \cellcolor[HTML]{FFFFFF}13.2                 \\
    lorem                       & \cellcolor[HTML]{9B9B9B}7.1                  & \cellcolor[HTML]{C0C0C0}235.2                & \cellcolor[HTML]{EFEFEF}6.4                  & \cellcolor[HTML]{FFFFFF}1.6                  & \cellcolor[HTML]{9B9B9B}5.1                  & \cellcolor[HTML]{C0C0C0}0.8                  & \cellcolor[HTML]{EFEFEF}54.6                 & \cellcolor[HTML]{FFFFFF}0.6                  & \cellcolor[HTML]{9B9B9B}7.0                  & \cellcolor[HTML]{C0C0C0}26.7                 & \cellcolor[HTML]{EFEFEF}6.0                  & \cellcolor[HTML]{FFFFFF}2.6                  & \cellcolor[HTML]{9B9B9B}7.3                  & \cellcolor[HTML]{C0C0C0}15.4                 & \cellcolor[HTML]{EFEFEF}4.6                  & \cellcolor[HTML]{FFFFFF}1.9                  \\
    sentiment-analysis          & \cellcolor[HTML]{9B9B9B}7.6                  & \cellcolor[HTML]{C0C0C0}593.0                & \cellcolor[HTML]{EFEFEF}6.4                  & \cellcolor[HTML]{FFFFFF}18.9                 & \cellcolor[HTML]{9B9B9B}5.3                  & \cellcolor[HTML]{C0C0C0}0.7                  & \cellcolor[HTML]{EFEFEF}1292.3               & \cellcolor[HTML]{FFFFFF}5.3                  & \cellcolor[HTML]{9B9B9B}8.0                  & \cellcolor[HTML]{C0C0C0}133.5                & \cellcolor[HTML]{EFEFEF}7.2                  & \cellcolor[HTML]{FFFFFF}20.7                 & \cellcolor[HTML]{9B9B9B}7.8                  & \cellcolor[HTML]{C0C0C0}108.0                & \cellcolor[HTML]{EFEFEF}7.1                  & \cellcolor[HTML]{FFFFFF}22.2                 \\
    thumbnail                   & \cellcolor[HTML]{9B9B9B}7.0                  & \cellcolor[HTML]{C0C0C0}312.6                & \cellcolor[HTML]{EFEFEF}5.3                  & \cellcolor[HTML]{FFFFFF}25.5                 & \cellcolor[HTML]{9B9B9B}4.7                  & \cellcolor[HTML]{C0C0C0}0.6                  & \cellcolor[HTML]{EFEFEF}157.8                & \cellcolor[HTML]{FFFFFF}1.3                  & \cellcolor[HTML]{9B9B9B}5.8                  & \cellcolor[HTML]{C0C0C0}43.1                 & \cellcolor[HTML]{EFEFEF}5.5                  & \cellcolor[HTML]{FFFFFF}13.1                 & \cellcolor[HTML]{9B9B9B}5.4                  & \cellcolor[HTML]{C0C0C0}31.4                 & \cellcolor[HTML]{EFEFEF}4.5                  & \cellcolor[HTML]{FFFFFF}9.8                 
    \end{tabular}
    \caption{Cold-start overhead breakdowns in milliseconds. The numbers are averages from the same 100 rounds as Figure~\ref{fig:microbench}.
    The breakdown follows our model
    $max(c, (\frac{pgs_{unique}\times P}{bw_{disk}})) + init + (pgs_{shared}\times lat_{mem})$.
    \textbf{A-D} stands for the model's four clauses from left to right. Latencies in D column are the difference between
    the measured latency and \regular's latency. \regular stands for
    booting a function normally from the kernel.}
    \label{table:overhead-breakdown}
\end{table*}

\paragraph{Cold-Start Function Execution Time.}\label{sec:eval-exec}

We measure cold-start function execution time from the moment the host sends a
request to the VM until the host receives a response.
Figure~\ref{fig:microbench-exec} shows execution latencies normalized to
\snapfaas.

With fewer copy-on-write page faults, \seuss has comparable or
faster execution latencies than \snapfaasnoprof and \snapfaas.  Specifically,
\snapfaasnoprof and \snapfaas are 16.4\% and 13.4\% slower, respectively,
than \seuss for the Go \texttt{sentiment-analysis} function.

With no shared pages across function instances, \reap avoids all copy-on-write
costs incurred by \snapfaasnoprof and \snapfaas during execution. For the minimal
Python3 \texttt{lorem} function, for example, \snapfaasnoprof is 68.9\% slower than
\reap and \snapfaasnoprof is 74.7\% slower \reap for Go function
\texttt{sentiment-analysis}. However, our results show that \reap's execution
latencies are unpredictable, particularly if \reap misses some pages used
during execution, which much be fetched on-demand from disk.

In general, longer-running functions, such as \texttt{ocr} in Java and
\texttt{alexa-reminder} in Node.js, see smaller variations in execution
latencies across the four snapshot designs.


\paragraph{Cold-Start End-to-End Latency.}\label{sec:eval-e2e}

The end-to-end latency measures from the start of VMM process to when the
host receives a response from the VM. This is the sum of cold-start boot
latency and cold-start execution time and is the most important latency metric
for FaaS workloads. Figure~\ref{fig:microbench-e2e} shows end-to-end
latencies normalized to \snapfaas.


Without the working set optimization, \snapfaasnoprof is already at least comparable to
the existing \seuss and \reap for all functions.
Specifically, \snapfaasnoprof is up to 10.1x as fast as \reap for Node.js function \texttt{lorem}
and up to 5.3x as fast as \seuss for Python3 function \texttt{sentiment-analysis}.

With the working set optimization, \snapfaas further improves the speed-up.
Specifically, \snapfaas is up to 1.8x as fast as \snapfaasnoprof for Go function
\texttt{sentiment-analysis}, is up to 9.9x as fast as \reap for Node.js function \texttt{lorem},
and is up to 6.1x as fast as \seuss.


For long-running functions, such as Java function \texttt{ocr}, all strategies
are similar since the execution time dominates. Figure~\ref{fig:speed-up-trend}
shows the trend that shorter functions experience higher speed-ups.
Additionally, Figure~\ref{fig:speed-up-trend} shows that \snapfaas and \snapfaasnoprof
are the closest to the optimal case when there is no cold-start overhead.

\subsection{Cold-Start Overhead Breakdown}
    
\begin{figure*}[t]
    \centering
    \includegraphics[width=\textwidth]{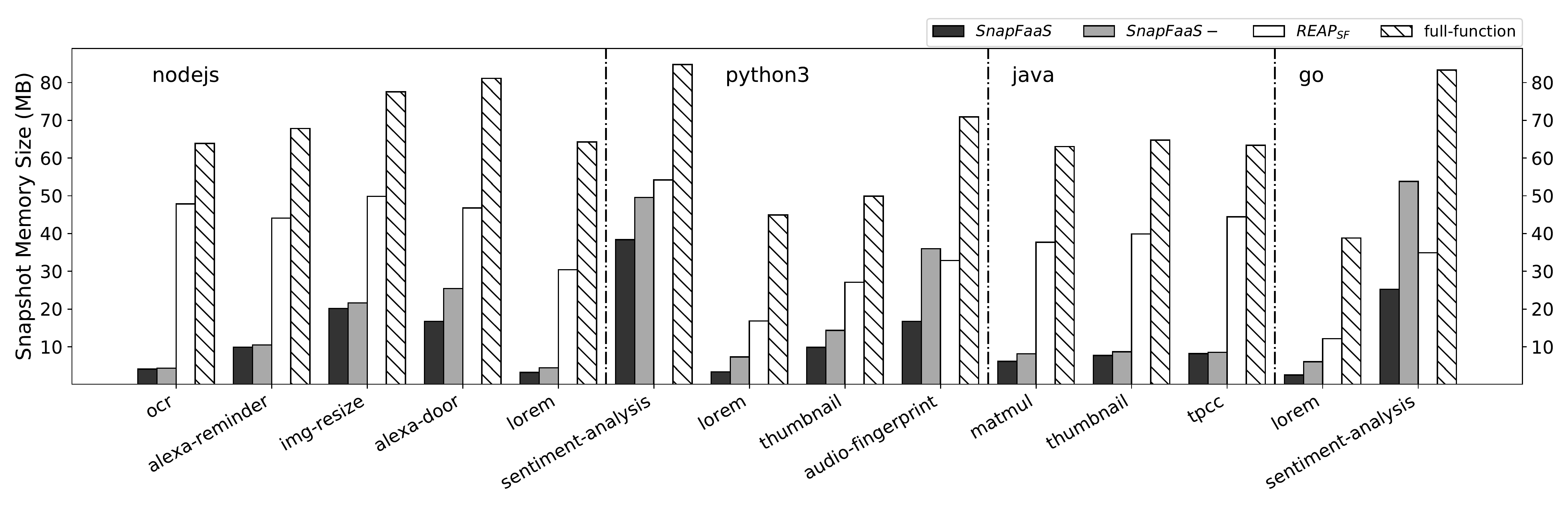}
    \caption{Sizes (MB) of memory eagerly restored from the disk. Compared with
    the sizes of full-function snapshots, \reap, \snapfaasnoprof, \snapfaas reduces
    the sizes of memory eagerly restored from the disk. \snapfaasnoprof achieves
    the reduction by caching common states with no working set approximation.}
    \label{fig:snapshot-sizes}
\end{figure*}

Table~\ref{table:overhead-breakdown} presents cold-start latency overhead breakdowns
of \snapfaasnoprof, \snapfaas, \seuss, and \reap.

Recall that we model cold start latency overhead as
\begin{equation*}
    max(c, (\frac{pgs_{unique}\times P}{bw_{disk}})) + init + (pgs_{shared}\times lat_{mem}).
\end{equation*}

Our empirical results show the follows where A to D stands for the model's four clauses from left to right.

\paragraph{A.} There is a constant overhead that the system
spends pre-configuring the VMM process and the VM and restoring non-memory states, e.g., CPU registers.

\paragraph{B.} For \snapfaasnoprof, \snapfaas, and \reap, there is a non-constant overhead that the system spends
restoring memory from the disk while this overhead for \seuss is constant and
small because \seuss only does \textbf{file-mmap}s to restore memory.

Figure~\ref{fig:snapshot-sizes} shows all evaluated functions' eagerly restored memory sizes
along with their full-function snapshot memory sizes. We can see that \snapfaasnoprof by caching common states
reduces more memory than \reap and that \snapfaas, with the working set optimization, can further
reduce memory sizes. The large memory size reductions
leads to reductions in cold-start boot time (Figure~\ref{fig:microbench-boot}) and results in
reductions in cold-start end-to-end latency (Figure~\ref{fig:microbench-e2e}).

Note that this restoration latency does depend on disk bandwidth utilization.
Our results do show various disk bandwidth utilizations intra and inter snapshot strategies.
For example, under \reap, Node.js function \texttt{alexa-door} only consumes 81 MB/s bandwidth on average
while Node.js function \texttt{alexa-reminder} consumes 100 MB/s bandwidth on average and
\texttt{alexa-door} under \snapfaasnoprof consumes 355 MB/s. Our implementation simply uses the \texttt{readv}
system call. Optimizations may be possible particularly for \reap. However, we want to point that
even if \reap can manage to consume the same bandwidth as \snapfaasnoprof, it still fundamentally loads
significantly more memory from disk into memory compared with \snapfaasnoprof.

\paragraph{C.} For \snapfaasnoprof, \snapfaas and \reap, there is a more or less constant overhead
that the system spends doing remaining initialization work, connecting to the host through VSOCK in this case.
For \seuss, in contrast, the system spends extra time importing the function from the source in addition
to connecting to the host.

\paragraph{D.} Empirically, there exists execution slow-down for all snapshot strategies even though
in theory \reap should observe no slow-downs because the inputs during evaluation are the same as during working set generation.
The results suggest that executions are not necessarily deterministic (e.g.\ languages' garbage collecting)
causing accesses to on-disk pages. For the rest three, \snapfaasnoprof and \snapfaas consistently experience
higher slow-down than \seuss. This follows from that, compared with \snapfaasnoprof and \snapfaas,
\seuss does more initialization work after restoring from the \base snapshot so that it experience
fewer copy-on-write faults during execution.

\subsection{Memory Overhead}\label{sec:eval-size}

\begin{figure}[t!]
    \centering
    \includegraphics[width=\linewidth]{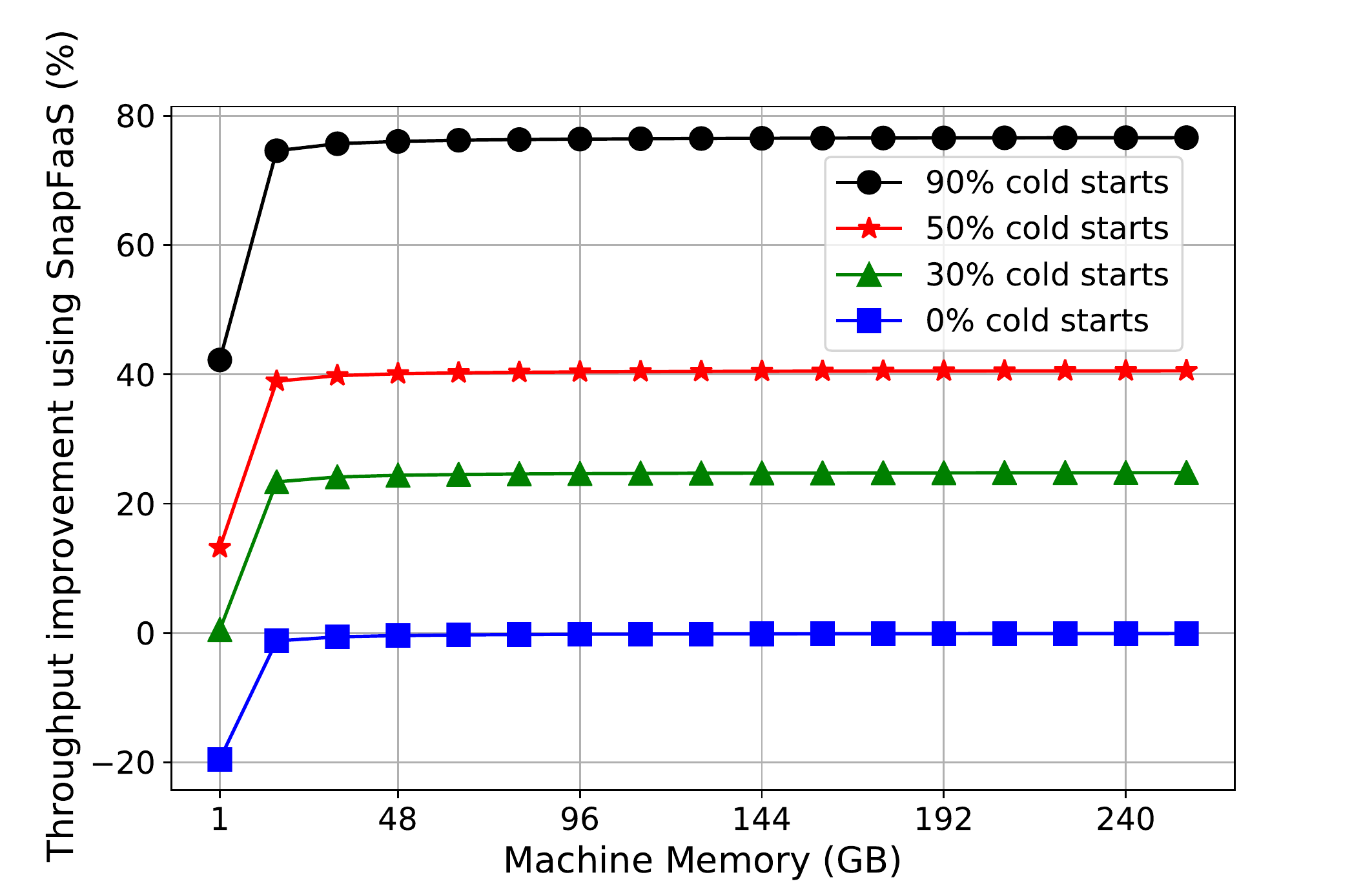}
    \caption{Throughput difference using \name{} vs \regular under 
    simulated workloads. When available memory is small and there no 
    cold-start requests, \name{}' memory overhead hurts throughput. However,
    when 30\% or more requests result in cold-starts, \name{} improves
    through
    25\%-77\%.}
    \label{fig:eval:throughput_simulation}
\end{figure}

\name{} requires each worker to have an in-memory copy of the \base snapshot for
each supported runtime. For the runtimes we implemented, this includes 60MB for
Node.JS, 40MB for Python, 60MB for Java, and 36MB for Go. In total, each
worker incurs a memory overhead of 196MB. On our experimental machines, this
amounts to 0.1\% and 0.3\% of the available 192GB and 64GB of memory,
respectively.

Less available memory means fewer functions can run concurrently on the same
machine. Conversely, faster cold-start latency means that throughput per
available slot is higher. How do these competing forces affect overall
throughput?

Figure~\ref{fig:eval:throughput_simulation} shows the throughput difference
using \name{} and regular VM initialization under a simulated workload with
varying proportions of requests resulting cold-starts. As expected, with no
cold starts (i.e.\ all requests are for recently invoked functions) \name{}
has lower throughput because it can run fewer VMs concurrently. However, with
as few as 30\% of requests resulting in cold starts, \name{} has 25\% higher
throughput. When most requests result cold-starts \name{} has over
75\% higher throughput than regular initialization.

\section{Related Work}\label{sec:related}

Prior work has looked into mitigating cold-start overhead with
checkpoint/restore techniques~\cite{snowflock, replayable-execution}.
Snowflock~\cite{snowflock} targets stateful applications in
traditional cloud computing environments. Their VM forks abstraction achieves
sub-second VM cloning. Replayable Execution~\cite{replayable-execution} is
recent work that target FaaS applications.
Replayable Execution uses checkpointing and demand paging to
boot a JVM environment in 54ms. Snapshots in Replayable Execution is taken
after JVM initialization and before loading user applications. Their JVM
initialization captures the maximal common state for their Java workload.
This approach is equivalent to our base snapshots.




Numerous research projects propose lightweight virtualization techniques.
Unikernels~\cite{osv, unikernel} reduce startup latency by minimizing the
guest OS based on applications and removing kernel-userspace isolation. No
FaaS systems currently use unikernels in production, but
SEUSS~\cite{seuss} shows that snapshots can be used with unikernels to further
reduce FaaS cold start latency. LightVM~\cite{lightvm} improves startup
latency by optimizing the Xen hypervisor. ukvm~\cite{ukvm} builds a
specialized virtual machine monitor on top of KVM for unikernels to reduce
startup latency. Solutions that improve hypervisor or VM monitor performance
can further benefit \name{} and are orthogonal to our approach.

Some CDN providers have begun using JavaScript and WebAssembly based sandboxes
to run FaaS-style
computations~\cite{cloudflare-workers,fastly-terrarium}. Some research
projects also explored using language sandboxes to run FaaS workloads.
Boucher et al. proposed using Rust's static types to isolate FaaS
computations~\cite{boucher}, and Splinter~\cite{splinter} uses a compile-time
sandbox based on Rust's static types to enable low-resource sandboxing of
computations running in a fast key-value store.

In general, language-based approaches offer orders of magnitude faster cold
start time and lower memory overhead compared with virtualization-based
sandboxes. The above systems all report cold start latencies on the order of
10s of microseconds. However, they trade generality: none of these approaches
can offer a full Linux environment --- Rust and JavaScript sandboxes in
particular only support applications written in those languages. Many
FaaS workloads rely on a variety of other languages as well as a wide
variety of existing code designed to run on Linux such as machine learning
libraries and image compression tools.

\section{Discussion}
Our results show that Linux VM-based snapshots have a lower-bound of
about 15 ms for our setup if the SSD peak bandwidth is achieved. While there is
an open space of snapshot designs, we argue that new designs cannot
significantly improve on our results without breaking the FaaS abstraction. This
has important implications for practitioners and researchers in this space.
Cold-start overheads limit the utilization of FaaS significantly when execution
times are very low. As a result, when targeting environments with low latency
requirements~\cite{fastly-terrarium,granular} system builders should avoid the
containerized Linux abstraction. Instead, FaaS systems that target high cluster
utilization and low latency must sacrifice the portability and flexibility of a
Linux interface in favor of language-specific or other limited APIs with better
fundamental performance characteristics.

\section{Limitations \& Future Work}\label{sec:limitations}

Using snapshots restoration to alleviate cold-start latency exacerbates two
important security concerns.
Instances spawned from the same snapshot share a large portion of their base
memory. This helps performance but renders attack (e.g.\ from a malicious
request payload) mitigation techniques based on address space layout
randomization (ASLR~\cite{aslr}) ineffective. Recent
work~\cite{shuffler,timely-rerand,clonewars} has shown that re-randomizing
memory and code can be done with reasonable overheads in many
cases~\cite{shuffler}. We intend to evaluate the use of re-randomization within
FaaS functions on \name{}.
Similarly, using copy-on-write shared pages for kernel and language runtime
memory introduces the potential for cache-based timing channels between
functions on the same machine~\cite{heyyou}.


Our current implementation of \name{} has some known limitations that we are
actively fixing. We currently do not capture the VSOCK device, the host-guest
channel, in either snapshots. As a result, even though the state in which the
channel ends right before any requests is predictable, in the current
implementation the system still initializes the channel through computation.
At the moment, each base snapshot only supports a particular VM memory size
For example, a Python3 base snapshot created on a VM with 128MB of memory cannot
be used to start a 256MB function. Supporting each VM size requires one base snapshot.
However, we believe this is fixable using well-known memory ballooning strategies
and intend to implement this functionality in \name{}.
Some runtimes (Node.js in particular) use the OS' random number generator
during initialization regardless of whether user code needs randomness.
Because there is very little entropy during sandbox initialization, the kernel
does not consider the randomness pool ready for a very long time, blocking
runtime initialization during snapshot generation. Our current
implementation manually insecurely adds ``randomness'' to the pool. However
we intend to incorporate the VirtRNG driver in the future. VirtRNG would allow the
guest VM to simply pre-seeded randomness from the host.

Substituting normal sandbox creation for efficient and fast snapshot-based
techniques as in \name{} opens a number of research directions we have not yet
explored in depth and leave for future work.
One example is that \name{} might help mask the performance and resource
overhead of heavier operating systems. Today, FaaS platforms use stripped down
operating system distributions to mitigate cold-start latency and reduce
per-function memory overhead. In a typical FaaS environment, without snapshots,
this makes sense.  However, the result typically lacks common developer
conveniences, such as DBus on Linux. Because \name{} loads the base snapshot
lazily we expect a more complete OS interface to have limited or no
per-function memory or performance overhead.


\section{Conclusions}\label{sec:conclusions}

We presented \name{}, a snapshot for FaaS system based on Linux VM.
We first think from first principles modeling
the fundamental overhead of snapshot restoration. Then, we have the model guide
our design leading to the \base-\diff split snapshot.
\name{} delivers near-optimal cold-start overhead with negligible memory
overhead.
\name{} and all of the experimental infrastructure is open source and available at https://fakeplatform.biz/mindyour/beeswax.

\bibliographystyle{plain}
\bibliography{references}

\end{document}